\documentstyle[12pt,axodraw]{article}
\begin{document}
\pagestyle{empty} \quad\vskip -1.5cm
\centerline{ hep-ph/0009292 \hfill          SLAC-PUB-8601
\footnote{Work supported by the Department of Energy, Contract 
DE-AC03-76SF00515}
}
%
\begin{center}{\bf \large
Large Two-Loop Contributions to $g-2$ \\
from a Generic Pseudoscalar Boson
}

\vskip 1cm

{\large
Darwin Chang$^{a,b}$, We-Fu Chang$^{a}$,\\
Chung-Hsien Chou$^{a}$,  and Wai-Yee Keung$^{c}$}\\

{\em $^a$NCTS and Physics Department, National Tsing-Hua University,\\
Hsinchu 30043, Taiwan, R.O.C.}\\

{\em $^b$Stanford Linear Accelerator Center, Stanford University, Stanford, 
     CA 94309, USA}\\

{\em $^c$Physics Department, University of Illinois at Chicago, IL
     60607-7059,  USA}\\

\end{center}
\date{\today}
\begin{abstract}
We calculate the dominant contributions to the muon $g-2$
at the two-loop level due to a generic pseudoscalar boson that may 
exist in any exotic Higgs sector 
in most extensions of the standard model.  
The leading effect comes from diagrams of the Barr-Zee type.
A sufficiently light pseudoscalar Higgs boson can give
rise to contribution as large as the electroweak contribution which
is measurable in the next round of $g-2$ experiment.  
Through the contribution we calculated here, the anticipated improved 
data in the recent future on the muon $g-2$ can put the best limit on the 
possible existence of a light pseudoscalar boson in physics beyond 
the standard model.

\end{abstract}
\vskip 0.5cm
\centerline{PACS numbers:  14.80.Cp,  14.60.Ef,  13.40.Em}
\vskip 0.5cm
\newpage
%
\pagestyle{plain}
%

Precision measurement of the anomalous magnetic moment of the
muon, $a_\mu\equiv {1\over2}(g_\mu-2)$, can provide not only a sensitive
test of quantum loop effects in the electroweak standard  model (SM),
but can also probe the effects of some potential ``new physics''. 
The experimental average in 1998 Particle Data Book gives\cite{PDG} 
$a_\mu^{\rm
exp} = 11659230 (84) \times 10^{-10}(\pm 7.2 {\rm\ ppm})$.  Recent
measurements by E821 experiment at Brookhaven give \cite{E821}
$a_\mu^{\rm exp} = 11659250 (150) \times 10^{-10}(\pm 13 {\rm\ ppm})$
(1997 data) and $a_\mu^{\rm exp} = 11659191 (59) \times
10^{-10}$ $(\pm 5 {\rm\ ppm})$ (1998 data).  

The E821 experiment has just announced the most accurate
result\cite{E821c}  from
the 1999 data sample. The weighted mean measurement  is
\begin{eqnarray}
a_\mu^{\rm exp} = 116592023 (151) \times 10^{-11} (\pm 1.3 {\rm\ ppm})\ .
\end{eqnarray}
Its high precision poses a direct challenge to the theoretical
prediction of  $a_\mu$, which are usually divided into
four sources,
\begin{eqnarray}
a_\mu=a_\mu^{\rm QED}+a_\mu^{\rm Hadronic}+a_\mu^{\rm EW}+\Delta a_\mu ,
\end{eqnarray}
representing QED, hadronic, electroweak and the exotic (beyond
the standard model) contributions respectively.  The QED loop
contributions have been computed to very high order \cite{ref:aqed-cl}
\begin{eqnarray}
a_\mu^{\rm QED} &=& 
{\alpha\over 2 \pi} +0.765857376(27) \left( {\alpha\over  \pi}\right)^2
+24.05050898(44) \left( {\alpha\over  \pi}\right)^3
\nonumber\\ &&
+ 126.07(41) \left( {\alpha\over  \pi}\right)^4
+ 930(170) \left( {\alpha\over  \pi}\right)^5   \ .
\label{eq:qed}
\end{eqnarray}
The most precise value for the fine structure constant
$\alpha= 1/137.03599958(52)$ can be obtained by
inverting the similar formula for the electron $g_e-2$ from the
data\cite{Kinoshita95}.  This gives
\begin{eqnarray}
a_\mu^{\rm QED} &=& 116584705.7(2.9)\times 10^{-11}  \ ,
\end{eqnarray}
with precision much higher than the expected experimental reach.  
The SM electroweak contribution up to two-loop level gives 
$a_\mu^{\rm EW}= 152(4)\times 10^{-11}$\cite{ckm,ckm2} for 
$\sin^2\theta_W = 0.224$ and $M_H = 250$ GeV  (In comparison, the
one-loop SM electroweak contribution is $195 \times 10^{-11}$).
The hadronic contribution\cite{hadron} due to  the hadronic vacuum
polarization diagram has the largest uncertainty from the strong interaction,
$a_\mu^{\rm Hadronic} = 6739(67)\times 10^{-11}$.
However, such uncertainty is still smaller than the experimental
error in $a_\mu$, even before  results from future planned experiments 
which intend to measure the hadronic vacuum polarization directly.
The total value in standard model is\cite{ref:cl},
\begin{equation}
a_\mu(\mbox{SM})=a_\mu^{\rm QED}+a_\mu^{\rm Hadronic}+a_\mu^{\rm EW}=
116591597(67)\times 10^{-11} (\pm 0.57 {\rm\ ppm}).
\end{equation}

The current measurement is well above the standard model prediction 
by a $2.6\sigma$ effect, $\Delta a_\mu=(42.6\pm 16.5) \times 10^{-10}$.
Given that $a_\mu^{\rm Hadronic}$ and $a_\mu^{\rm EW}$
are both positive, one can conclude that the current data have
already probed these contributions, but they are not enough to fit the data.
Extra contributions from new physics is needed.
At 90\%CL, $\Delta a_\mu$ lies\cite{ref:cl}  between 
$(+21.5$ --- $+63.7)\times 10^{-10}$.
Note that the negative $a_\mu^{\rm EW}$(two-loop) 
plays an important role in claiming this discrepancy.

Even without the recent experimental improvement, $g-2$ data has
already provided non-trivial constraints\cite{constraints} on
physics beyond the standard model. For example, the constraint on 
the minimal supersymmetric standard model (MSSM) due to its 
one-loop contribution to $g-2$, via
smuon-smuon-neutralino and
chargino-chargino-sneutrino loops, is well known\cite{gm2mssm}. 
The resulting constraint
depends on the masses of supersymmetric particles and $\tan\beta$.

In theories beyond the standard model, there are usually
additional scalar or pseudoscalar bosons.  
In particular, some of the pseudoscalar bosons can potentially be light 
because of its pseudo-Goldstone nature, accidentally or otherwise.
However, in collider search, it is known that 
searching  for  the pseudoscalar neutral boson is much harder than 
the scalar neutral or charged one.  
Therefore it is particularly interesting to see if one can constrain 
the pseudoscalar boson using low energy precision experiments.
In this paper we wish to report that if the extended theory  has a light 
enough pseudoscalar boson, its two-loop contribution to muon $g-2$ can
be as large as the one-loop electroweak effect. As a result the muon
$g-2$ can provide a very strong probe on a large class of
theories beyond the standard model.  

The one-loop contributions to $g-2$ from a scalar or pseudoscalar boson 
have been presented  many times in the literature\cite{oneloop}.  
Besides two powers of
$m_\mu$ that is required by kinematics and definitions, the one-loop 
contribution is further suppressed by another two powers of $m_\mu/M_a$.
However, the result is enhanced by a logarithmic loop factor, 
$\ln m_\mu/M_a$, coming from
the diagram in which the photon is emitted by the internal muon.  
Therefore for a light enough Higgs mass, some limit can be derived from 
$g-2$ data just based on one-loop result.  
Nevertheless, as we shall see later, 
the two loop contribution is typically larger 
than the one-loop one by a factor of 2--10 for Higgs mass from 10 -- 100 GeV.
In addition, the one-loop and two-loop contributions have different signs for
both scalar and pseudoscalar contributions.  
Therefore the one-loop contribution
actually partially cancels the larger the two-loop contribution.
 
The two-loop contribution of a scalar boson has been calculated in 
Ref.\cite{ckm,ckm2} in the context of the standard model.  The 
contribution of any scalar boson beyond the standard model can in 
principle be extracted from that calculation and we shall not dwell 
on this here except to note that the scalar boson gives negative
contribution while the pseudoscalar gives positive contribution
to $\Delta a_{\mu}$.  
Also, we have parameterized our input Lagrangian as model independent as
possible in order to make our gauge invariant result widely applicable to 
a large class of models.

For Higgs mass larger than roughly 3 GeV,
the dominant Higgs related contribution to the muon
anomalous magnetic moment is through the two-loop Barr-Zee type
diagram\cite{bz}, as in Fig.~1. Compared with the one-loop graph, the Yukawa
coupling of the heavy fermion $f$ in the inner loop together with the
mass insertion of the heavy fermion in the two-loop graph will give rise to
$(m_f/m_\mu)^2$ enhancement which can overcome the extra loop
suppression factor of $\alpha/16\pi^2$.  
The internal gauge boson can be a photon or a $Z^0$.  The $Z^0$
contribution is typically smaller by two order of magnitude.
It is included in this manuscript just for completeness.
Note that unless CP violation occurs in the Higgs potential, there is 
no two-loop Barr-Zee type contribution to $g_\mu-2$ associated with 
pseudoscalar boson and an inner gauge boson loop.
\begin{center}
\begin{picture}(200,160)(0,0)
\Oval(100,80)(30,50)(0) \Photon(100,110)(100,160){5}{5}
\Text(110,150)[lc]{$\downarrow$ $\gamma$ $(k,\mu)$}
\DashArrowLine(20,20)(60,60){5} \Photon(140,60)(180,20){5}{5}
\Text(35,40)[rc]{$a^0(p)$} \Text(80,45)[cc]{$f$}

\Line(97,53)(103,47) \Line(97,47)(103,53)

\Text(105,60)[cc]{$l$} \Text(130,115)[lc]{$f$}
\Text(140,90)[rc]{$l+q$} \LongArrow(135,102)(140,98)
\Text(60,90)[lc]{$l+p$}
\LongArrow(60,98)(65,102 ) 
\LongArrow(85,51 )(80,53)  
\LongArrow(121,53 )(117,51)  
\LongArrow(160,55)(170,45) 

\Text(175,50)[lc]{$\gamma$, $Z^0$ $(q,\nu)$}
\ArrowLine(0,20)(20,20) \Text(10,10)[cc]{$\ell$}
\ArrowLine(20,20)(180,20) \Text(100,10)[cc]{$\ell(-q)$}
\ArrowLine(180,20)(200,20) \Text(190,10)[cc]{$\ell$}
\end{picture}
\end{center}
{\small Fig.~1  The dominant two-loop graph involving a pseudoscalar boson
that contributes to $g_\ell-2$.  The cross location denotes a possible
mass insertion.}

\noindent
The form of the gauge invariant vertex function 
$\Gamma^{\mu\nu}$ of a pseudoscalar boson $a^0$ of momentum 
($p$) turning into two photons $(-k,\mu)$, $(q,\nu)$ due to the
internal fermion or gauge boson loop is
\begin{equation}
\Gamma^{\mu\nu} = 
   P(q^2) \epsilon^{\mu\nu\alpha\beta} p_\alpha q_\beta  \ .
\end{equation}
In general, the heavy fermion generation dominates in the loop. 
The Yukawa coupling is parameterized  in a model independent
expression,
\begin{equation} 
{\cal L}= i \frac{g A_f m_f}{2 M_W} \bar{f}\gamma_5 f a^0, 
\end{equation} 

Integrating the fermion loop momentum, we obtain the form factor
\begin{equation}
P(q^2)= N_c^f{ g A_f e^2 q_f^2 m_f^2 \over 8\pi^2 M_W}
\int^1_0   {dz \over m_f^2-z(1-z) q^2 }  \ ,
\end{equation}
where $m_f$ and $q_f$ are the mass and the charge of the internal
fermion  in the loop. The color trace gives $N_c^b=N_c^t=3$, $N_c^\tau=1$.
The above vertex is further connected to the lepton
propagator to produce anomalous magnetic dipole moment 
$a_\ell^{\gamma a^0}$ for the lepton $\ell$,
\begin{equation}
a_\ell^{\gamma a^0}=\frac{\alpha^2 }{8\pi^2\sin^2\theta_W}
\frac{m_\ell^2 A_\ell}{M_W^2}\sum_{f=t, b,\tau } N_c^f q_f^2 A_f 
{m_f^2 \over M_a^2} {\cal F }\left({m_f^2 \over M_a^2}\right) \ ,
\label{eq:aP}
\end{equation}
\begin{equation}
{\cal F }(x)= \int^1_0 { \ln\frac{x}{z(1-z)} dz  \over x-z(1-z)} \ .
\end{equation}
${\cal F}(1)= \frac{4}{\sqrt{3}}\mbox{Cl}_2(\frac{\pi}{3})$,
with the Clausen's function
$\mbox{Cl}_2(\theta) =
-\int^\theta_0\ln \left(2\sin\frac{\theta}{2}\right) d\theta$.
As $ x \gg 1 $, $x{\cal F}(x)$ has the
asymptotic form $2+\ln x$. On the other extreme limit
$x \ll 1$, ${\cal F}(x)$
approaches to $ \frac{\pi^2}{3}+ \ln^2 x  $. Our result is consistent 
with that from an unphysical Higgs boson in SM\cite{ckm2}. 

For the graph with the inner photon replaced by $Z^0$ boson,  its
contribution to $a_\mu$ can be calculated in a similar fashion,
\begin{equation}
a_\ell^{Z^0 a^0} = {  \alpha^2  m_\ell^2 A_\ell g^\ell_V 
\over 8\pi^2\sin^4\theta_W\cos^4\theta_W  M_Z^2}
\sum_{f=t,b,\tau}{N_c^f A_f q_f g^f_V  m_f^2 \over M_Z^2 - M_a^2}
 \left[ 
 {\cal F}\left( \frac{m_f^2}{M_Z^2}\right)- 
 {\cal F}\left( \frac{m_f^2}{M_a^2}\right) 
 \right]  \  , \label{eq:aPZ}
\end{equation}
with $g_V^f={1\over2} T_3(f_L)-q_f\sin^2\theta_W$.  Note that, for both
pseudoscalar and scalar boson contributions, only the vector
coupling of $Z^0$ to heavy fermion contributes to the effective
vertex due to Furry theorem.  Numerically, this $Z^0$ mediated
contribution turns out to be about two order of magnitude smaller
than that of the photon mediated one. One suppression factor comes
from the massive $Z^0$ propagator and the other one comes from 
the smallness of the leptonic vector coupling of $Z^0$ boson, 
which is proportional to $(-{1\over4} + \sin^2\theta_W)\sim-0.02$.

Taking the pattern of Yukawa couplings in MSSM as an example, we set
$A_f$ as $\cot\beta$  ($\tan\beta$) for the $u$ (or $d$)-type fermion.  The
contributions due to top quark $t$ , bottom quark $b$ and tau lepton $\tau$ 
in the loop respectively as well as the total are displayed in Fig.~2 
for both $\tan\beta = 30$ and $50$.  
In this MSSM pattern  the $t$ contribution is insensitive to
$\tan\beta$.   In addition, both the $b$ and the $\tau$ contributions,
which are roughly the same order of magnitude,
dominate over that of the top quark one for large enough $\tan\beta$ and
light enough pseudoscalar mass $M_a$.
For $M_a \stackrel{<}{\sim}  15$ GeV, 
the $\tau$ contribution is larger than the 
$b$-quark contribution.
The total two-loop photonic  contribution from the pseudoscalar boson,
$a_\mu^{\gamma a^0}$, can be as large as $10^{-8}$ for 
a large $\tan\beta$ when $M_a \leq 10$ GeV as shown in Fig.~2.  
For example, for $M_a=10$ GeV and $\tan\beta=50$,
$a_\mu^{\gamma a^0}$(2-loop) $=1.2 \times 10^{-8}$, 
which is above the upper limit allowed by the current experiment bound.
Generically, for $M_a\sim $ 80--100 GeV, $\tan\beta\sim 50$, $a_\mu$
ranges in (7 -- 9)$\times10^{-10}$ which is close to the electroweak
contribution.  Note that the pseudoscalar contribution has the same
sign as the hadronic or electroweak contributions.
%
To derive constraint from the data one must combine the one- and two-loop 
contributions.
The well-known one-loop contribution\cite{oneloop} due to the pseudoscalar
$a^0$ is 
\begin{equation}
a_{\ell}^{a}\hbox{(1-loop)} =
- {{m_\ell}^2 \over 8{\pi}^2 {M_a}^2} 
  \left(\frac{g A_\ell m_\ell}{2 M_W}\right)^2 
H\left({m_\ell^2 \over M_a^2}\right) \ ,\quad \hbox{with }
H(y)=     \int_0^1 {x^3 dx \over 1-x+x^2 y } \ .
\end{equation}
For small $y$, $H(y)\to -\ln y - {11\over 6}  > 0 $. 
Note that the one-loop contribution is always negative in contrast 
to the two-loop contribution.  
In Fig.~2, we draw the absolute value of the one-loop contribution
for easy comparison.
For small $M_a$ such as $10$ GeV, the one-loop contribution can be as large 
as half of the two loop contribution and produce cancelling effect 
in $g_\mu-2$.
Complete cancellation occurs around 3 GeV for large $\tan\beta$.
However the one-loop effect becomes smaller for larger $M_a$ 
due to its additional suppression factor of 
$(m_\mu^2/M_a^2) \ln (M^2_a/m^2_\mu)$ and basically negligible for 
$M_a > 50$ GeV.

The up-to-dated measurement from E821 (1999) has already 
indicated 2.6 $\sigma$ deviation from the standard model. 
The allowed contribution from new physics falls into a small interval,
$\Delta a_\mu$ between 
$(+21.5$ --- $+63.7)\times 10^{-10}$ at 90\%CL. 
The positive two-loop contribution is able to fit the data for large 
$\tan\beta \sim 50 $ and $M_a \stackrel{<}{\sim} 40$ GeV, as
illustrated in Fig.~2.
Note that for $M_a$ lighter than roughly 3 GeV, the negative one-loop 
contribution dominates and gives the overall negative $\Delta a_\mu$,
which is disfavored by the current E821 data.

In CP conserving MSSM, there is a lower bound\cite{mamssm,expmamssm}
on $M_a \geq 88$ GeV, which is only based on partial scanning with
certain choices of benchmarks in the MSSM.
Furthermore,
in more general supersymmetric models or in general two or more Higgs
doublet models\cite{ma}, very little can be said about the potentially
light pseudoscalar Higgs boson.  Experimental
constraint\cite{ref:expa} on $M_a$ from LEP data is correlated to a
rather light scalar Higgs boson.  
The model independent nature of our calculation makes it possible to
derive useful information of  the pseudoscalar boson sector in any
theory beyond the standard model using the hard earned data on muon
$g-2$.  

Note that in general multi-Higgs doublet models, the $\tan\beta$
factor in our analysis may be supplemented by additional factor of
mixing matrix elements.
In addition, in any specific model, there may be additional two-loop
contributions, such as the ones involving the physical charged Higgs
boson or the neutral scalar boson.  We assume that these contributions
do not accidentally cancel each other.  Given that the experimental
limit on the masses of the charged Higgs boson as well as the neutral
scalar boson are already quite high, it is very unlikely they will
cancel the contribution of a relatively light pseudoscalar boson.

In conclusion, in this letter we report a set of analytic formulas for
the two-loop contributions of a generic pseudoscalar boson to lepton
anomalous magnetic moment.  Such pseudoscalar bosons may exist in any
theory beyond the standard model and they are typically harder to
constrain using collider experimental data. In this paper, we show
that strong constraint on such sector can be derived from the
precision data on muon anomalous magnetic moment from the going and
future experiments.  We hope our work add importance and urgency to
these low energy precision experiments.

\noindent {\bf Acknowledgments} This work was supported in
parts by National Science Council of R.O.C., by U.S. Department of
Energy (Grant No. DE-FG02-84ER40173).    
We wish to thank A. Pilaftsis, M. Krawczyk, T. Rizzo and 
Bogdan Grzadkowski for discussions and 
DC wishes to thank SLAC Theory Group for hospitality.

\noindent
{\bf Note added:} 
While the original manuscript of this work was in the review process, 
the E821 experiment announced its new measured value. Based on this
non-trivial result, we have updated our Fig.~2, which shows the
implication on the mass of  the pseudoscalar Higgs boson.

\newpage
\epsffile[90 90 506 580]{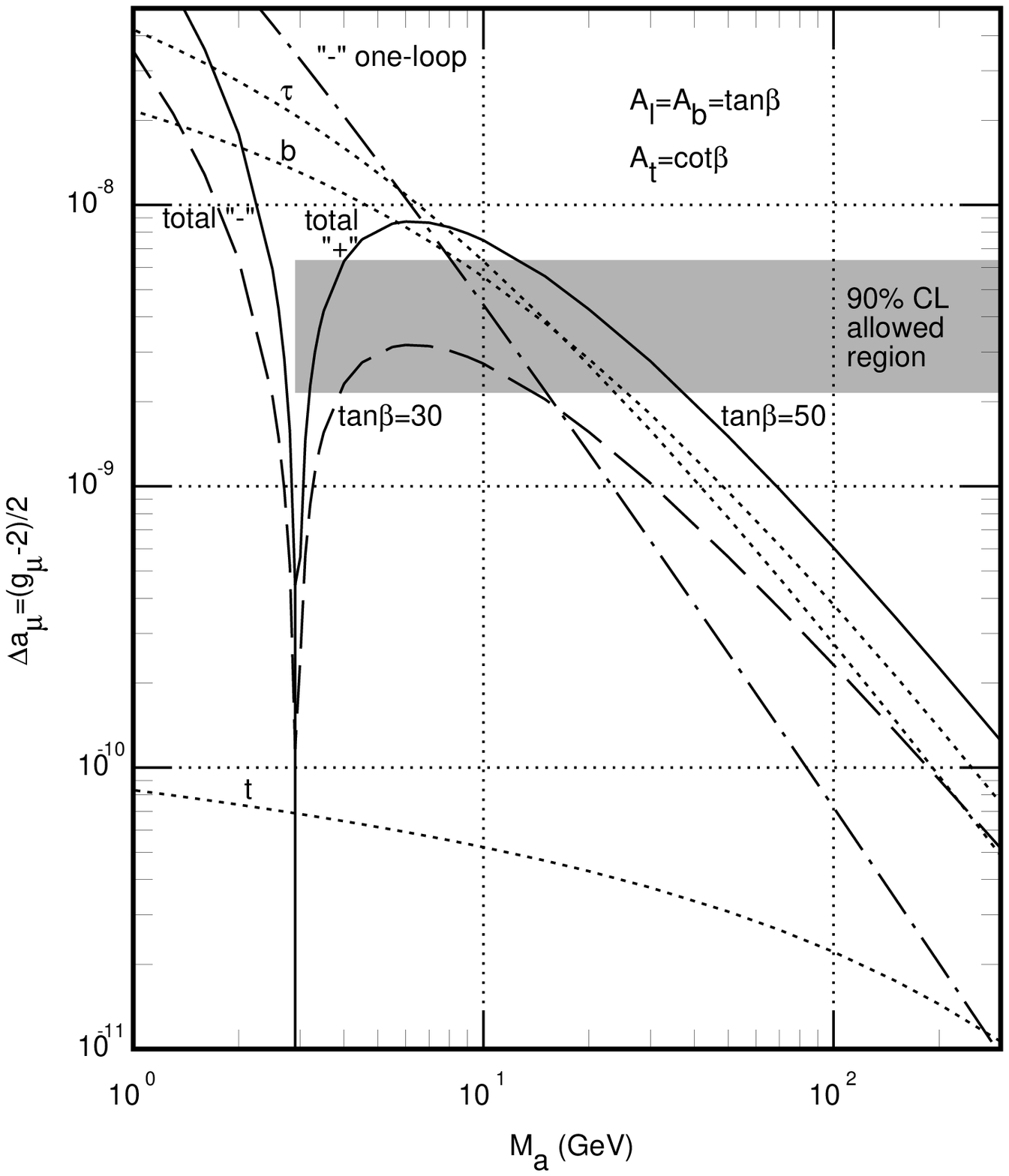}
\noindent
\vfill
{\small Fig.~2: 
The dotted lines plot the 
positive two-loop contributions from the inner 
$t$,$b$, $\tau$ loops to $g_\mu-2$ due to the pseudoscalar $a^0$ versus $M_a$ 
at $\tan\beta=50$, while the dashed-dotted line plots
the negative contribution from the one-loop diagram.
The sum in solid (dashed) curve shows cancellation at low $M_a$
mass for $\tan\beta=50$ (30).
The shaded areas shows the allowed contribution to $\Delta a_\mu$ 
from  new physics  at 90\%CL based on E821 measurement (1999). 
The region of negative $\Delta a_\mu$, mainly from the one-loop  
contribution  for a very light $M_a< 2.9$ GeV, 
is not favored by data.}
\end{document}